\documentstyle[twocolumn,aps,epsfig]{revtex}
\draft

\begin{document}
\title{The origin of the highest energy cosmic rays \\
Do all roads lead back to Virgo?
}

\author{Eun-Joo Ahn$^1$, Gustavo Medina-Tanco$^2$,
Peter L. Biermann$^{3}$, and Todor Stanev$^4$
}
\address {
$^1$Department of Physics,
Seoul National University, Seoul, Korea}
\address {     
$^2$Institute of Astronomy and Geophysics,
University of Sao Paulo, Sao Paulo, Brasil}
\address {
$^3$Max Planck Institut f{\"u}r Radioastronomie,
Bonn, Germany}
\address {
$^4$Bartol Research Institute,
University of Delaware, Newark, DE, USA
}


\wideabs{
\maketitle
\begin{abstract}
\widetext

Introducing a simple Galactic wind model patterned after the solar wind we show
that back-tracing the orbits of the highest energy cosmic events suggests that
they may all come from the Virgo cluster, and so probably from the active
radio galaxy M87.  This confirms a long standing prediction, as well as a
theoretical model for the radio galaxy jet emission of M87.  With this picture
in hand, one clear expectation is that those powerful radio galaxies that have
their relativistic jets stuck in the interstellar medium of the host galaxy,
such as 3C147, will yield limits on the production of any new kind of particle,
expected in some extensions of the standard model in particle physics.
\end{abstract}
}

\narrowtext


\section{Introduction}

 The origin of the highest energy particles observed in the universe
 continues to present a major enigma to physics.  These particles
 reach energies as high as $3 \, 10^{20}$ eV.  The flux of such
 nuclei is expected to drop sharply at $5 \, 10^{19}$ eV due to the
 interaction with the microwave background, commonly referred to as the
 GZK-cutoff after its discoverers \cite{GZK}.  However, the number of
 particles known to be beyond $10^{20}$ eV continues to increase, with now 14
 published, and a further 10 expected from new observations with  HIRES
 and a reanalysis of the Yakutsk data~\cite{new_data}.

 There are three basic difficulties:  First, we need to find a site
 that can either accelerate particles to such energies, or
 produce them outright through the decay of even higher energy
 particles (topological defects for example,
 \cite{Bhatta99}).  While there are
 many sites which are in principle capable to produce such particles
 \cite{Hillas84,JGCR}, there is only one class that has
 been argued to require protons at such energies in the source, namely radio
 galaxies with powerful jets and/or hot spots, such as M87 \cite{BS87}
 or Cyg A.  M87 has been under suspicion to be the primary source for
 ultra high energy cosmic rays for a long time \cite{GS63,HP80}.
 For both compact and extended radiojets as well as hot spots such as
 in M87  the emission of synchrotron
 radiation~\cite{BS87} with a steep cutoff at frequencies
 about $3 \, 10^{14}$ Hz
 implies an initial turbulence injection scale of the Larmor radius
 of protons at $10^{21}$ eV.
 Radio galaxies of sufficient power can provide through acceleration
 sufficient momentum to particles to overcome losses within the space
 limitation and the source magnetic and radiation fields~\cite{Hillas84,BS87}.
 Gamma ray bursts are another possible class.

 Secondly, there is the additional difficulty of getting these particles
 to us, that is, overcoming the losses in the bath of the cosmological
 microwave background \cite{GZK}.  That implies that the source should not be
 very much further than 20 Mpc, and that corresponds to the travel distance for
 the presumably charged particles, not necessarily as short as the light travel
 time distance.  There are only two certain candidates, the radio galaxy
 M87 at about 20 Mpc, and the radio galaxy NGC315
 at about 80 Mpc - this latter distance is hard to accept already.
 Hence, we require that not much additional travel time should be spent
 through scattering in large scale magnetic irregularities
 \cite{Ryu98}.

 The third difficulty is the explanation of the nearly isotropic
 distribution of the arrival directions of these events.
 The anisotropy expressed in a weak correlation with the supergalactic
 plane, or the detection of multiples mostly along the supergalactic plane
 \cite{Stanev95,Agasa_sgp} is a small effect.
 The arrival directions are best described as isotropic, at least in
 the accessible sky in the North.
 There are two extreme explanations of the isotropy, one is to argue
 that we have many sources contributing, even at the highest particle
 energies, and the other is that we have one source dominating at the
 maximum energies, and the particle orbits are bent.  In the second
 option the bending should not add substantially to the travel time.

 There are several ways out of these difficulties.
 The highest energy particles could be particles generated by
 topological defects~\cite{Bhatta99}
 or from extensions of supersymmetry~\cite{FB98}.
 Alternatively, they could be cosmologically
 local~\cite{Berezinsky98,Weiler99}.
 If many sources contribute to the observed particle fluxes
 they have to be fairly common in space, as in the scenarios
 of accelerating such particles in the environment of quiescent
 black holes~\cite{BG99}, or gamma ray bursts~\cite{Waxman97}.

 Here\footnote{inquiries to plbiermann@mpifr-bonn.mpg.de}
 we present the consequences of introducing a
 magnetic Galactic wind in analogy to the solar wind.
 The magnetic field of the wind bends the
 particle orbits without adding substantial travel time.

\section{A model for a magnetic galactic wind}

 It has long been expected that our Galaxy has a wind~\cite{SF76}
 akin to the solar wind~\cite{Parker58}.  Recent modelling~\cite{Galwind99}
 shows that such winds can be quite fast, and ubiquituous.

 It seems plausible that this wind is powered by the combined action
 of cosmic rays and magnetic fields~\cite{Galwind99}
 and starts in the hot phase of the interstellar medium
 seen in X-rays by ROSAT.
 This phase has a density of $3 \, 10^{-3}$ particles per cc,
 a temperature of about $4 \, 10^6$ K, a radial scale of about 5 kpc,
 and an exponential scale in $z$, perpendicular to the disk, of almost
 2 kpc \cite{ROSAT97}.  We note that the corresponding equipartition
 magnetic field strength is 10 microGauss.

 Parker has shown~\cite{Parker58} that in a spherical wind the poloidal
 component of the magnetic field becomes negligible rather quickly with
 radius $r$, the radial component decays with $1/r^2$, but the azimuthal
 part of the magnetic field  quickly becomes dominant with
 $B_{\phi} \, \sim \, \sin \theta /r$ in polar coordinates.

 An available measure of magnetic field along any line of sight is
 the Rotation Measure:  It is proportional to the
 line of sight integral of the product of electron density and magnetic
 field component (including the sign) from us to a distant linearly
 polarized radio source. We verified that
 the magnetic field topology of the Parker model is consistent
 with the data~\cite{Kronberg94} for a base density  below that of
 the hot interstellar medium.

 Cosmic ray driving is similar to radiation driving of winds in massive
 stars, and so \cite{SB97} magnetic fields can lead to an increase of
 the momentum of the wind.
 For a steady wind and Parker topology the Alfv{\'e}n speed in the wind
 becomes independent of radius $r$ for large distances.
 The ultimate wind velocity is then a
 small multiple of the Alfv{\'e}n velocity, suggesting
 a rather strong magnetic field.  On this basis we have adopted
 here a magnetic field somewhat higher than in other current models
 \cite{Galwind99}.

 The data on the sign of the azimuthal component show that in the
 disk of the Galaxy there are reversals but in the direction
 of the anti-center, the part of the sky most relevant for calculating
 orbits of energetic charged particles, the field points to the
 direction of galactic longitude about 90 degrees
 \cite{Simard,Kronberg94}.  That
 means immediately that positively charged particles traced
 backwards have their origin above us, at high positive galactic latitudes.

 There is one additional observational argument in favor of such
 a topology of the magnetic field:  Krause \& Beck \cite{Krause98}
 have found that the dominant magnetic field symmetry in spiral
 galaxies points approximately along the local spiral arms inwards
 towards the center.  This effect is unexplained at present,
 but it seems nevertheless to be observationally well established.
 One might expect that magnetic field sign reversals in the disk
 should translate into similar reversals in the wind.  However, if
 the  dominant directionality is established  by the most active
 spirals arms, then the symmetry as found by Krause \& Beck
 should prevail well outside the disk.  We will assume this to be
 the case. This symmetry implies an overall current in the Galaxy
 possibly driven by star formation through stellar winds and
 cosmic rays.

 If the entire disk has this symmetry, then there cannot be a sign
 reversal of the azimuthal component close to the disk along the
 direction perpendicular to the disk.  This implies that the radial
 magnetic field component $B_r \sim 1/r^2$ is pointed inwards
 inside the disk and in its immediate neighborhood below and above.
 However, considering all of $4 \pi$ the radial component has to
 be pointed outwards roughly over half the sky, and pointed inwards
 in the other half in order to satisfy the condition of a source
 free magnetic field.
 Thus, one  possible and simple configuration is that $B_r$ is pointed
 inwards within  30 degrees of the disk both above and below the disk,
 and pointed outwards within 60 degrees of both poles.  Such a pattern
 is different from the Solar wind, where the components change sign in
 mid-plane.

  We adopt the simplest possible model.  We assume
 that the magnetic field in the galactic wind has a dominant azimuthal
 component, and ignore all other components.  We assume that this azimuthal
 component has the same sign everywhere.  This then means that the strength
 of this azimuthal component at the location of the Sun is the first key
 parameter, and the distance to which this wind extends is the second one.
 Since the irregularities in magnetic fields should decay into the
 halo,  the total magnetic field strength near the Sun may be used as an
 approximation to the local halo wind field, and in fact one might
 expect a lower limit.
 Most measures of magnetic field underestimate its strength.
 Therefore we will consider for reference a model which has a field
 strength near the Sun of 7 microGauss~\cite{Beck96}.  The second
 parameter, the distance to which this wind extends, is more uncertain:
 Our Galaxy dominates its near environment well past our neighbor, M31,
 the Andromeda galaxy, and might well extend its sphere of influence to
 half  way to M81.  Therefore we will adopt as outer the halo wind radius
 half the distance to M81, 1.5 Mpc.

\section{Tracing the path backwards}

 To follow the particle trajectories in the Galactic halo we trace
 protons backwards~\cite{Stanev97}
 from their arrival direction at Earth.
 We use the 14 published cosmic ray events above 10$^{20}$ eV,
 the list from Watson (included in \cite{LB99}) and the new list
 from AGASA \cite{AGASA99}. There is a big uncertainty with
 the energy estimate of the highest energy
 Yakutsk~\cite{new_data} event,
 which we therefore exclude from the present analysis, and hence
 we arrive at a final tally of 13 events used.

\begin{figure}[!hbt]
\centerline{\psfig{figure=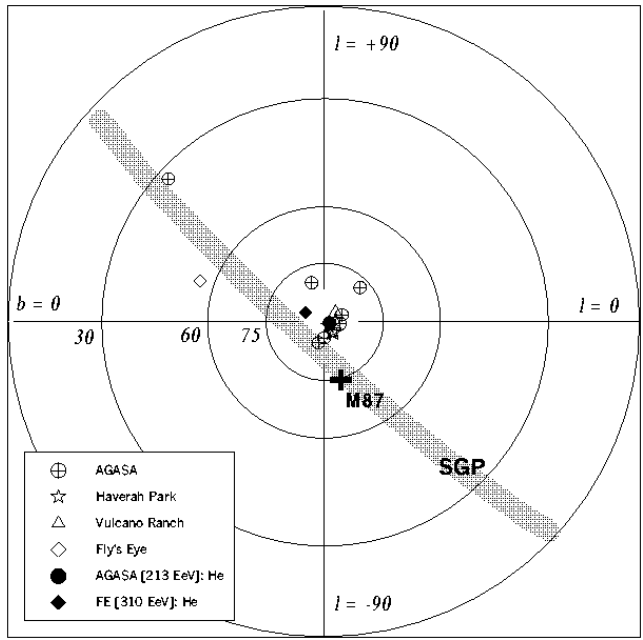,width=8.5cm}}
\vspace*{0.2truein}
\caption{ Directions in polar projection of the 13 highest
 cosmic ray events when they leave the halo of our Galaxy.
\label{plb_fig1}}
\end{figure}

 Fig.~\ref{plb_fig1} shows the directions of the events at that
 point when they leave the halo wind of our Galaxy in a polar
 projection.  For reference we show the direction to the active
 galaxy M87 (Virgo A), the dominant radio galaxy in the Virgo cluster.
 We show the two highest energy events twice: under the assumption,
 (i) that they are protons, and (ii) that they are Helium-nuclei
 (filled black symbols).  We also show the supergalactic plane
 as a shaded band.

 The interesting result of these model calculations is
 that the directions of all tracks point North.
 With the exception of the two events having the highest energy,
 all other 11 events can be traced to within less than about
 20 degrees from Virgo A.  Considering the
 uncertainty of the actual magnetic field distribution,
 we find then that all events are consistent with arising originally from
 Virgo A. Since these particles are assumed to be accelerated
 out of cosmic gas, about 1/10 of all particles may be
 Helium nuclei with the same energy per particle.
 If the two highest energy events are in fact He nuclei, all 13 events
 point within 20$^\circ$ of Virgo A.

 If Virgo A is indeed the acceleration site of the highest energy
 cosmic ray events, they all require systematic bending at a ten to
 twenty degree level. Such bending could be easily accomodated within
 the plausible magnetic field strength within the supergalactic sheet
 from here to Virgo~\cite{Vallee,Ryu98}. Bending by 20 degrees on a
 pathlength of 20 Mpc  implies a regular transverse  magnetic
 field strength of order 2 nanoGauss for 10$^{20}$ eV protons.
 In fact, the expected total magnetic field strength in the supergalactic
 plane is significantly higher~\cite{Vallee,Kronberg94,Ryu98}.

How critical are our assumptions for these results?\\[1truemm]
$\bullet$ The assumption of the symmetry of the magnetic field
 above and below the Galactic disk is important; if
the magnetic field were to have opposite sign above and below the disk, then
all those events that arrive below the disk, would in fact go down and point in
the general direction of the Galactic South pole.\\[1truemm]
$\bullet$ The value of the magnetic field, here adopted as
 7 microGauss, for the wind near the Sun, is a key parameter.
 A decrease of this value by up to $\sim$ 30\% would not change
 essentially the results.  If the magnetic
 field were considerably weaker, however, of order one microGauss,
 then the effective pointing to a common direction of origin
 would be largely removed.  However, it would still be sufficient
 to smear arrival directions, especially if its structure is
 regular.\\[1truemm]
$\bullet$ The scale of the Galactic wind here 1.5 Mpc, is not
 a critical parameter, since the calculations show that most of the
 bending happens within the first few 100 kpc.

\section{Discussion and implications}

This particular model can be tested in several ways:

 It has been suggested repeatedly in the literature, that radio galaxies
 may inject a major fraction of the total energy output (of order 0.1)
 in energetic particles~\cite{GS63,UHECRII,Ensslin97,Leahy99}.
 In order to determine this fraction for M87 we need a model for the
 magnetic field structure along the
 supergalactic sheet, and test it with orbit calculations.

 Second, if the appearance of pairs and triplets of events
 \cite{Agasa_sgp} along the supergalactic sheet is confirmed,
 then clearly the magnetic field structure along the sheet may
 be the reason in that it gives rise to some caustics with an enhancement
 just along the sheet direction.  Again, orbit
 calculations will be able to test this.

 Third, the very concept of a magnetic wind, driven by cosmic rays,
 but with an initial magnetic field as strong as in the disk,
 needs to be examined more closely.
 The conditions to be met are a) the available energetics,
 b) agreement with the Rotation Measure data,
 c) the condition that the wind be super-Alfv{\'e}nic, and
 d) that the wind does extend to large scales.  The model for Wolf Rayet star
 winds \cite{SB97} gives us some confidence, that this is indeed reasonable.
 Since this should be true for any galaxy with a high star formation rate
 and therefore, presumably with a high rate of cosmic ray production,
 this latter condition may be the easiest to test with sensitive
 absorption line data at high redshift, with observations of
 the shell of gas around these extended winds.

 Fourth, the current experimental statistics cover only a fraction
 of the Southern sky. We would expect some asymmetry between
 the Northern and Southern hemispheres of our sky.  How much asymmetry
 depends again on the more detailed magnetic field configurations,
 but the data from the Auger observatory will clearly provide stringent
 conditions on this as on any other model.

Fifth, if the model proposed here could be confirmed, then
it would constitute strong evidence that all powerful
radiogalaxies produce high energy cosmic rays, and that they do
this at a level of $\simeq$ 0.1 of their total power
output.  This then implies that compact radio galaxies \cite{FB98}
do provide a good test bed for particle interactions,
 since they have a large
screen of interstellar gas around the radio hot spots and jets as seen in
mm-wavelength radio data \cite{Papa99}.
  There the paradigm is materialized of
having a gigantic accelerator, and a beam dump.  These radio galaxies may be
used for particle interaction experiments in the sky. If a significant
 correlation in arrival direction between ultra high energy cosmic rays
 and this specific class of radio quasars could be confirmed~\cite{FB98},
 then properties of new particles could be constrained.

 In summary, we propose here that a very simple model for a
 Galactic wind rather analoguous to the Solar wind, may allow
 particle orbits at $10^{20}$ eV to be bent sufficiently to
 allow ``super-GZK'' particles to get here from M87,
 and also explains the apparent isotropy in arrival directions.\\[2truemm]
 {\bf Acknowledgments.}
This work was started at the 1999 astrophysics summer school
at the Vatican Observatory; both EJA and PLB are
grateful to George Coyne, S.J. and his colleagues for inviting us and for
the fruitful and inspiring atmosphere at these schools.
 PLB, GMT and TST would
like to acknowledge also useful discussions with Rainer Beck,
 Pasquale Blasi,
 E. Boldt, Glennys Farrar,
 J.L. Han, Anatoly Ivanov, Randy Jokipii, Frank Jones, Hyesung Kang,
 Phil Kronberg, Martin
Lemoine, Friedrich Meyer, Hinrich Meyer, Biman Nath,
Dongsu Ryu, Michael Salomon, G{\"u}nther Sigl, Alan
Watson, Arnold Wolfendale and many others.
 We thank Phil Kronberg for a careful reading of the manuscript.
Work with PLB is partially supported by a DESY-grant,
GMT is partially supported by the Brazilian agencies
FAPESP and CNPq, TST is supported by NASA grant NAG5--7009,
 and PLB and TST have jointly a grant from NATO.

\end{document}